\documentclass[twocolumn,amsmath,amssymb,10pt,superscriptaddress,a4paper,letterpaper,fleqn]{revtex4-1}
\usepackage{amssymb}
\usepackage{epsfig}
\usepackage{graphicx}
\usepackage{dcolumn}
\usepackage{array}
\usepackage{bm}
\usepackage{fancyheadings}
\usepackage{longtable}
\usepackage{multirow}
\usepackage{float}
\usepackage{afterpage}  
\usepackage{color}

\setlongtables
\usepackage[breaklinks=true,linkbordercolor={1 1 1}]{hyperref}

\begin{document}
\setcounter{page}{1}
\title{
\qquad \\ \qquad \\ \qquad \\  \qquad \\  \qquad \\ \qquad \\ 
Nuclear Science References  Database}
\author{
B. Pritychenko
}
\email[Corresponding author, electronic address:\\ ]{pritychenko@bnl.gov}
\affiliation{
National Nuclear Data Center, Brookhaven National Laboratory, Upton, NY 11973-5000, USA
}
\author{
E. B\v{e}t\'{a}k
}
\affiliation{
Institute of Physics, Slovak Academy of Sciences, 84511 Bratislava, Slovakia
}
\author{
B. Singh
}
\affiliation{
Department of Physics \& Astronomy, McMaster University, Hamilton, Ontario, Canada L8S 4M1
}
\author{
J. Totans
}
\affiliation{
National Nuclear Data Center, Brookhaven National Laboratory, Upton, NY 11973-5000, USA
}

\date{\today} 

\begin{abstract}
{
The Nuclear Science References (NSR) database together with its associated Web interface, is the world's only comprehensive source of easily accessible low- and intermediate-energy nuclear physics bibliographic information for more than 210,000 articles since the beginning of nuclear science. The weekly-updated NSR database provides essential support for nuclear data evaluation, compilation and research activities. The principles of the database and Web application development and maintenance are described. Examples of nuclear structure, reaction and decay applications are specifically included.  The complete NSR database is freely available at the websites of the     \linebreak National Nuclear Data Center  http://www.nndc.bnl.gov/nsr and the  \linebreak International Atomic Energy Agency http://www-nds.iaea.org/nsr.
}
\end{abstract}
\maketitle


\lhead{ND 2013 Article $\dots$}
\chead{NUCLEAR DATA SHEETS}
\rhead{B. Pritychenko, E. B\v{e}t\'{a}k  $\dots$}
\lfoot{}
\rfoot{}
\renewcommand{\footrulewidth}{0.4pt}

\section{Introduction}
\label{sec:Intro}

The NSR database is a bibliography of nuclear physics articles, indexed according to 
content and spanning from 1896 to the present day. The database originated at the Nuclear Data Project at Oak
Ridge National Laboratory as part of the systematic evaluation of
nuclear structure data \cite{ew78}, and was later adopted by the 
wider research community.  It has been used since the early 1960s to produce
bibliographic citations for nuclear structure and decay data evaluations published in Nuclear Data Sheets. 
 
 In October 1980, database maintenance and updating became the responsibility of the National Nuclear Data Center (NNDC) at 
Brookhaven National Laboratory (BNL).  The database has subsequently been through significant expansion, several 
modernizations, and technical improvements \cite{11Pri,wi05,pr06,pri12},   although 
the basic structure and contents have remained unchanged. Presently, NSR database compilations and Web developments have been conducted in 
collaboration with  the Nuclear Data Group at McMaster University, Canada and  Nuclear Data Section, IAEA \cite{04Zer}, respectively. The collaborative approach helped to improve the database content and develop new features.

 In this paper, we present the recent changes to NSR contents and features which make the database an essential nuclear bibliographic source. A brief description of the database, Web interface, and update policies are  given in the following sections.

\section{Database: Scope and Structure}
\label{sec:Scope}
The NSR database aims to provide primary and secondary bibliographic information for low- and intermediate-energy nuclear physics \cite{ra96}.
The diverse contents of the database are cataloged under seven major physics topics
\begin{center}
\begin{tabular}{p{5cm}p{5cm}}
{\sc Atomic Masses} & {\sc Nuclear Reactions} \\ [0.2cm]
{\sc Atomic Physics} & {\sc Nuclear Structure} \\ [0.2cm]
{\sc Compilation}   & {\sc Radioactivity}  \\ [0.2cm]
{\sc Nuclear Moments} \\
\end{tabular}
\end{center}

NSR entries include extensive information, starting with a unique eight-character identifier (NSR keynumber), 
journal/reference, publication year, article title, author list, journal digital object identifier (DOI) link, and a keyworded 
abstract (for articles reporting on appropriate physical quantities). All entries are stored in a relational database structure.

\section{NSR Keywords}
\label{sec:Keywords}
The main goal of NSR is to provide bookmarks for experimental and theoretical articles in nuclear science 
using keywords. Keywords serve a dual purpose in NSR 

\begin{itemize}
\item They are used to generate database {\it selectors}, which produce the correct article indexing and allow 
specific and detailed searches to be made quickly and easily (searching can also be done within the general text of entries).
\item They allow a user to  determine quickly which articles are of specific interest from a list of entries
returned following a given query.
\end{itemize}


By the very nature of the NSR database, the keyworded abstracts are very well structured. 
They begin with the topic identifier, as listed in Section \ref{sec:Scope}, and a list of nuclei, 
nuclear reactions, or decays follow. Then the measured and/or calculated/analyzed quantities are given, 
followed by deduced (derived) quantities. 

Measured quantities in NSR are based on the direct results of 
online measurements. For example, these primary quantities will 
include $\gamma$-transition energy and intensity, particle-$\gamma$ coincidences, etc. 
Other quantities, such as $\sigma$, S-factors, log {\it ft}, T$_{1/2}$ and B($\lambda$) values that are often derived offline, using 
the primary data, are considered deduced quantities. 
The same philosophy applies for calculated and analyzed quantities.


\section{NSR Web Retrievals}
\label{sec:Retrievals}
\begin{figure}
\begin{center}
\fbox{\includegraphics[height=6cm]{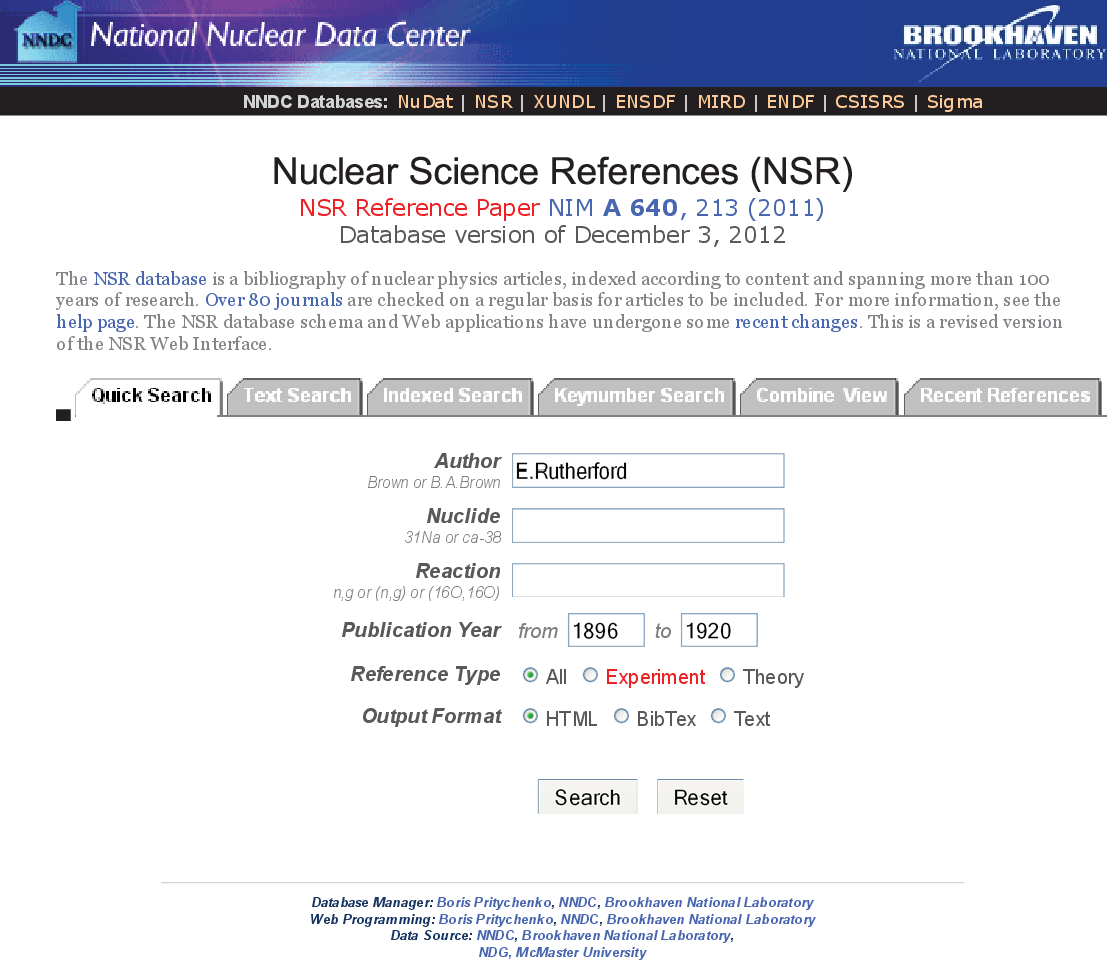}}
\caption{NSR Web Interface  http://www.nndc.bnl.gov/nsr. Example of Boolean search for author E.Rutherford and 1896-1920 time range.}
\label{fig3}
\end{center}
\end{figure}

The NSR Web Retrieval Interface is an integral part of both the NNDC and IAEA Web Services \cite{pr06,pri12,04Zer}. 
The Web interface is based on current Java technologies and provides retrievals of the database content in HTML, Text, BibTex, XML and PDF formats. 
As shown in Fig. \ref{fig3}, the main Web interface  consists of six sub-interfaces

\begin{center}
\begin{tabular}{p{5cm}p{5cm}}
{\sc Quick Search} & {\sc Text Search} \\ [0.2cm]
{\sc Indexed Search} & {\sc Keynumber Search} \\ [0.2cm]
{\sc Combine View}   & {\sc Recent References}  \\ [0.2cm]
\end{tabular}
\end{center}

The Quick Search allows a quick look-up of references for a given author, nuclide, or reaction within a publication period. 
The Text Search allows plain text searching of the title and keyword fields, whilst an Indexed Search allows a 
Boolean {\sc and} search over several indexed categories (e.g. author, nuclide, {\it etc.}). 
Keynumber Search retrieves the information for a specific article(s) given the NSR keynumber(s).
This type of specific retrieval is in large demand by nuclear structure evaluators. 
Finally, Combine View provides analysis and combination opportunities for previous retrievals, 
whilst Recent References provides downloads of quarterly compilation collections in text and PDF formats. 

\begin{figure}
\begin{center}
\includegraphics[height=6cm]{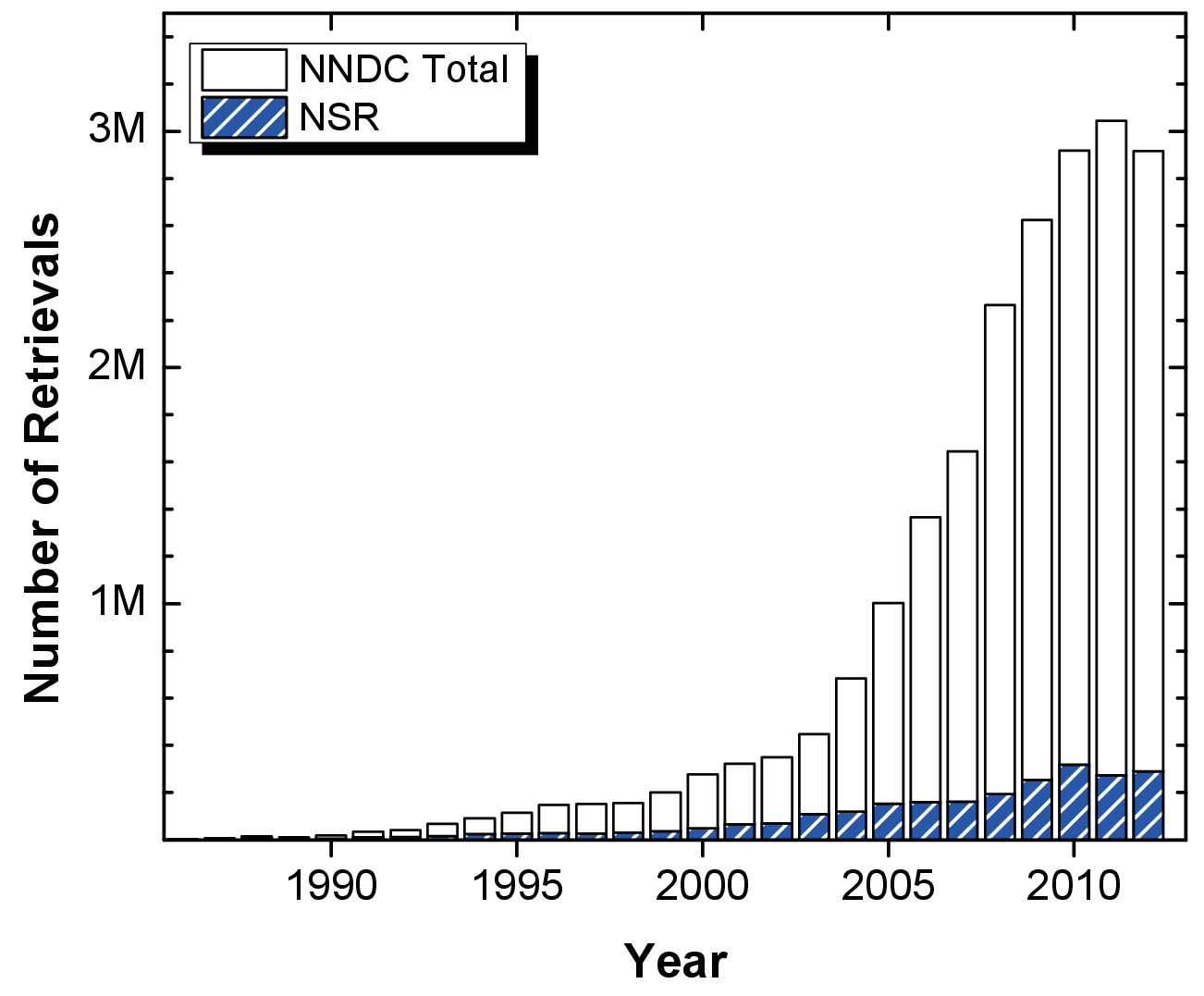}
\caption{The time evolution of the number of electronic retrievals - web, telnet and FTP - from 1986 to 2012.}
\label{fig4}
\end{center}
\end{figure}

An important part of monitoring NSR operation is a correct estimate of the database usage. NSR retrieval statistics are very conservative 
and  based completely on a count of successful database retrievals - any Web browser hits are ignored. 
The time evolution for NSR retrievals at NNDC over the last 25 years is shown in Fig. \ref{fig4}. 

\section{NSR Applications} 
\label{sec:Applications}
\begin{figure}
\begin{center}
\includegraphics[height=6cm]{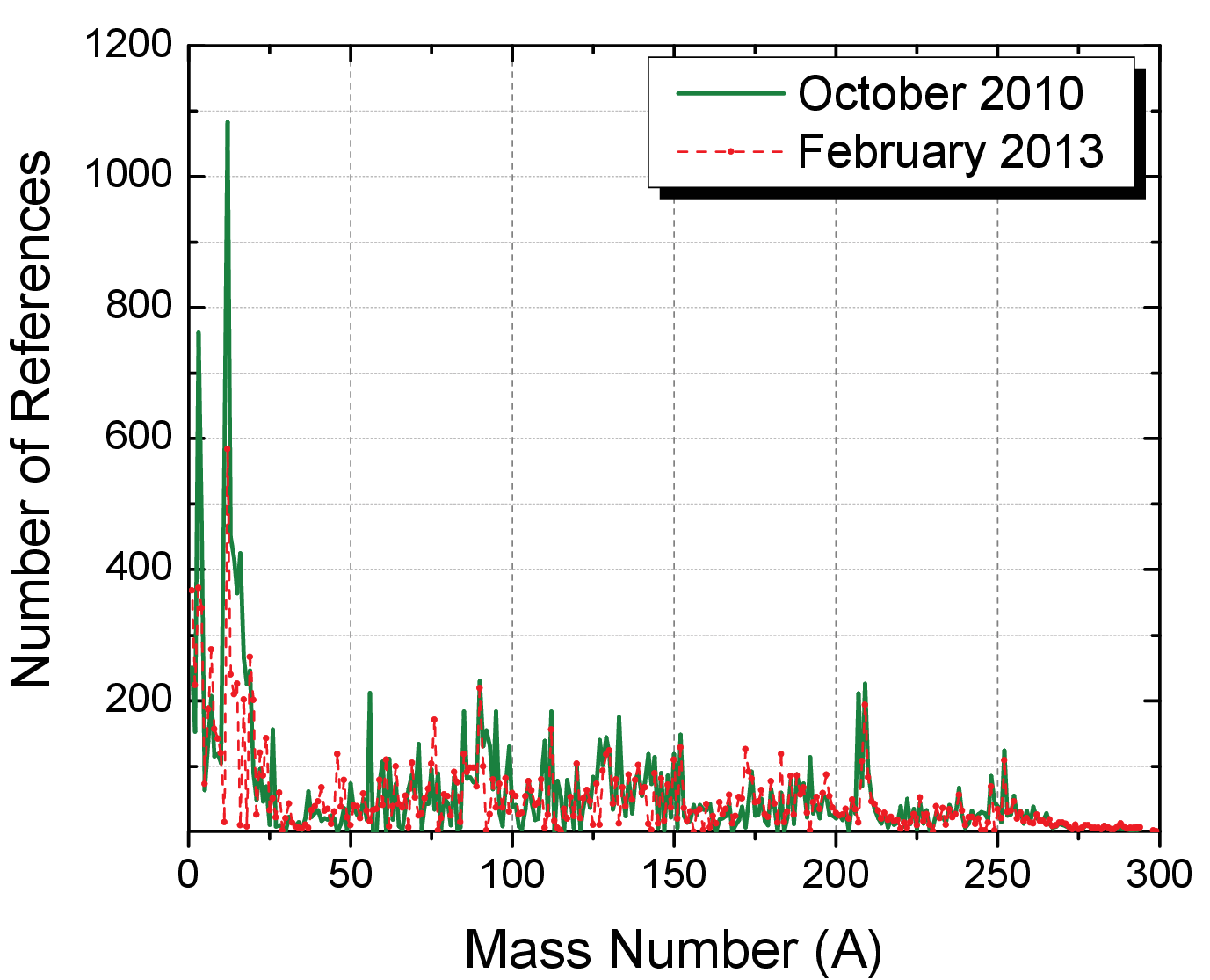}
\caption{Mass number distributions for references unevaluated by the NSDD network, as of October 2010 and February 2013 are shown in green and red colors, respectively.}
\label{fig6}
\end{center}
\end{figure}

The NSR database was initially created to support the Evaluated Nuclear Structure Data File (ENSDF) mass chain evaluations \cite{ensdf} . All references in ENSDF 
evaluations are specified by their NSR keynumbers.  Regular NSR database 
updates serve as an indicator for the international Network of Nuclear Structure and Decay Data Evaluators (NSDD) \cite{nsdd} on the requirement to
revisit a particular isobaric mass chain. 
As an example, Fig. \ref{fig6} shows number of references as a function of mass number, which have not yet been included in the ENSDF database. 
 As of February 2013, the average number of new references per mass chain is 52.

In addition to ENSDF mass chain or vertical evaluations \cite{ensdf,fi96}, NSR is actively used in a large number of horizontal 
evaluations  of atomic masses and NUBASE  \cite{12wan,12aud}, B(E2) and $\beta$$\beta$-decay values \cite{12pri,13pri} and compilation of the discovery of individual isotopes \cite{11Tho}.  
The NSR database and Web interface are also linked to a large number of other nuclear databases: ENSDF \cite{ensdf}, XUNDL \cite{xundl}, and EXFOR \cite{04Zer,exfor}. 
Thus, when a particular reference forms part of a compilation or evaluation in one of these other databases, NSR will provide a direct Web 
link to the publication.


\section{Conclusion and Outlook}
\label{sec:Conclusions}
The NSR database and  Web interface  provide transparent and easy access to nuclear  physics bibliographic information with direct links to the original articles and data, where possible. This project is conducted under the auspices of the U.S. Nuclear Data Program in a collaborative manner.

Recent additions include extension of NSR coverage from 1896 to 1911 and more targeted coverage of fundamental physics; over 600 articles of practical importance to nuclear science have been included. Further addition will improve database completeness by cross checking against the following sources
\begin{itemize} 
\item Decay Data Evaluation Project  references  \cite{ddep}.
\item EXFOR  database \cite{exfor}; until the 1990s, the scope of NSR was limited to nuclear structure physics, and therefore approximately 40$\%$ of EXFOR references are missing from NSR.
\item Discovery of Isotopes Project   references \cite{11Tho}.
\end{itemize}

Many features have been developed for nuclear scientists and specifically {\it reaction} data users, such as user-friendly Web retrievals, Web 
integration with the EXFOR database and improvements in NSR terminology/keywording. As a result, NSR has greater potential application in modern physics, 
as the major nuclear database that allows searches for rare isotope beam reactions.

{\it Acknowledgments:}
\label{sec:Acknowledgements}
We are grateful to M. Herman (BNL)  for his constant support of this project, to  D.F. Winchell (XSB, Inc.) and V. Zerkin (IAEA) for significant technical contributions,  
to  J. Choquette (McMaster University) for useful suggestions, and to M. Blennau (BNL) for carefully reading  the manuscript.  
This work was sponsored in part by the Office of Nuclear Physics, Office of Science of the U.S. 
Department of Energy under Contract No. DE-AC02-98CH10886 with Brookhaven Science Associates, LLC.





\bibliographystyle{model1a-num-names}
\bibliography{<your-bib-database>}



\end{document}